\documentclass[preprint]{aastex}
\usepackage{emulateapj5, psfig, epsfig}
\slugcomment{accepted by The Astronomical Journal - May 10, 2002}

\shorttitle{NGC~5634: a globular cluster in the Sgr stream?}
\shortauthors{Bellazzini et al.}

\def\kpc{{\rm\,kpc}}
\def\Gyr{{\rm\,Gyr}}
\def\kms{{\rm\,km\,s^{-1}}}

\begin{document}

\title{The stellar population of NGC~5634. \\
    A globular cluster in the Sagittarius dSph stream?
    \thanks{Based on observations made with the Italian Telescopio Nazionale
     Galileo (TNG) operated on the island of La Palma by the Centro
  Galileo Galilei of the CNAA (Consorzio Nazionale per l'Astronomia e
 l'Astrofisica) at the Spanish Observatorio del Roque de los Muchachos
                  of the Instituto de Astrofisica de Canarias.}}

\author{Michele Bellazzini, Francesco R. Ferraro}
\affil{INAF - Osservatorio Astronomico di Bologna, Via Ranzani 1, 40127, 
Bologna, ITALY}
\email{bellazzini@bo.astro.it, ferraro@apache.bo.astro.it}

\author{Rodrigo Ibata}
\affil{Observatoire de Strasbourg, 67000 Strasbourg, France}
\email{ibata@newb6.u-strasbg.fr}


\begin{abstract}

We present  the first $(V,V-I)$  Color -- Magnitude Diagram  (CMD) for
the poorly studied  globular cluster NGC~5634.  The CMD  shows a steep
Red  Giant Branch  (RGB) and  a predominantly  blue  Horizontal Branch
(HB): both these characteristics suggest  a low metal content for this
cluster. From the position of the RGB in the CMD we estimate $[Fe/H] =
-1.94 \pm  0.10$ and $E(B-V)=0.06  \pm 0.01$.  The CMD  presented here
reaches $V\simeq 23$,  allowing us to obtain the  first measure of the
Main  Sequence   Turn-Off  (TO)  $V_{TO}=  21.22\pm   0.15$  for  this
cluster. By combining this figure  with the apparent luminosity of the
{\it Zero Age HB} (ZAHB), $V_{ZAHB}=17.90 \pm 0.10 $ we obtain $\Delta
V^{HB}_{TO}=3.32\pm0.16$, a value which  is fully compatible with that
derived for  the bulk  of Galactic globulars.   We also derive  a true
distance  modulus of  $(m-M)_0 =  17.17 \pm  0.12$ corresponding  to a
distance  of $\sim  27.2 \kpc$.   Most interestingly,  the  cluster is
shown to have  position and radial velocity fully  compatible with the
orbit  of the  Sgr  dwarf spheroidal  (dSph)  galaxy.  The  similarity
between the  stellar populations of  the cluster and the  Sgr globular
Ter~8 and the metal poor population of the Sgr dSph, also suggest that
NGC~5634 was  a former member of  this disrupting galaxy,  torn off by
the Galactic tidal field and now lost in the Sgr stream.
\end{abstract}



{}{}
\section{Introduction}

NGC~5634   is   a       high-latitude   globular   cluster
($l^{II}=342.21$,  $b^{II}=+49.26 $)  in the  northern  hemisphere. Its
luminosity \citep[$M_V=-7.66$,  according to][hereafter D93]{dj93} and
concentration ($c =  log (r_t/r_c) = 1.60$, where  $r_t$ and $r_c$ are
the  tidal radius  and  the  core radius,  respectively;  see D93  and
references  therein)  are  typical  figures for  a  Galactic  Globular
Cluster (GGC).

Despite its brightness,  the cluster was really poorly  studied in the
past, since no  Color - Magnitude Diagram (CMD)  has been published so
far. A possible cause of this,  could be the presence of a very bright
foreground star located  at only 1.7 arcmin to the  East of the center
of  the cluster.   This  observational drawback  may  have turned  the
attention of observers toward less problematic alternative clusters.

In  the \citet{pet86}  compilation of  GGC CMDs  only three  items are
quoted  for  this  cluster:  (1)  the search  for  variable  stars  by
\citet{ba45}, in which seven  RR Lyrae variable stars were identified;
(2) the follow-up study by \citet{lisa}, who assessed the light curves
and periods of  six of the seven variables found by  Baade; and (3) an
unpublished CMD by Racine, quoted  also by \citet{lisa}. All the above
works were based on photographic plates.

The  ADS database (as  well as  other databases)  does not  report any
further study to  date. NGC~5634 is included in  the {\em HST Snapshot
Survey  of  globular clusters}  \citep[GO-7470,  see][for details  and
references]{manu99}  collecting  HST-WFPC2  shallow photometry  for  a
large sample  of Galactic globulars.  However,  up to now,  only a few
parameters (such as  the HB level and the  RGB-bump) have been derived
from the  HST data-set and published  by this team \citep{manu99,bono};
while, the CMD of the cluster and its overall characteristics have not
been discussed in any paper, yet.

Globular  clusters are  crucial tracers  of the  early history  of the
Galaxy, in particular  the rare Outer Halo clusters  that are found at
Galactocentric distances $R_{GC} >  10$ kpc. Since NGC~5634 belongs to
this latter group, it is certainly worth of studying.

In  this  paper we  present  the {\it  first  CMD  of NGC~5634}.   New
estimates of reddening, metallicity  and distance are obtained and the
Horizontal  Branch (HB)  morphology is  discussed. A  differential age
estimate is also obtained by  comparison with well studied clusters of
similar metal content. The possible connection with the huge stellar 
stream left by the disrupting Sagittarius dwarf spheroidal galaxy along 
its orbital path \cite[Sgr Stream, see][and references therein]{sdss}
is also discussed.

The plan of  the paper is as follows. We  present the observations and
the data  analysis in  Section 2, the  overall properties of  the main
branches of the  CMD (RGB, HB) are presented  in Section
3.  Section 4 is devoted to the determination of the distance and the age
of the cluster.  Finally Section 5 is dedicated  to discuss the possible
connection between NGC5634 and the Sgr galaxy.

\section{Observations and Data Reduction}

\subsection{Observations}

The  data were obtained  at the  3.52 $m$  Italian telescope  {\it
Telescopio  Nazionale  Galileo} (TNG, Roque  de  los  Muchachos, La  Palma,
Canary Islands, Spain),  using the LSR-Dolores Camera  equipped with a
$2048 \times 2048$ pixels thinned and back-illuminated Loral CCD array
(gain$=0.97 e^-/ADU$, read-out noise  $9.0$ ADU rms).  The pixel scale
is $0.275$  arcsec/px, thus the total  field of view of  the camera is
$9.4  \times 9.4  ~arcmin^2$.  Two  fields have  been observed  in the
region of  the cluster:  (1) a field  roughly centered on  the cluster
center (Central Field; hereafter CF)  and (2) a second one centered at
$\sim 6$ arcmin NE with respect to the cluster center (Field NE).  The
observations were carried out  during one photometric night (March 19,
2001), under  not excellent seeing conditions ($FWHM\simeq  1.2 - 1.4$
arcsec) as a backup program
while the main targets were not visible. 

The data consists of two V and  two I $30 ~s$ exposures for the FC and
only one long exposure ($600 ~s$) in each band for the outer Field NE.
  
The  observational   strategy  was   similar  to  that   described  in
\citet{mb01}: the short exposures in the central part samples
the  majority of the  cluster Red Giants  and HB
stars, conversely, with the long exposures in the outer region the Main 
Sequence (MS)  stars are observed in a field less plagued by the
stellar  crowding.   The  two  fields   (FC  and  Field  NE)  have  an
overlapping region $\sim  3 \times 3$ arcmin wide,  which is used (see
Section 2.2) to derive robust coordinate and magnitude transformations
between the two fields.

\subsection{Data Analysis}

The  raw  images were  corrected  for bias  and  flat  field, and  the
overscan  region  was  trimmed  using standard  IRAF\footnote{IRAF  is
distributed by  the National  Optical Astronomy Observatory,  which is
operated by the Association of Universities for Research in Astronomy,
Inc.,   under  cooperative   agreement  with   the   National  Science
Foundation.} procedures. Each couple  of $t_{exp}=30~s$ images, in each
band, has been  averaged, so that the final  analysis was performed on
each averaged V and I image.

The PSF-fitting procedure was performed  independently on each V and I
average images, using a version of DoPhot \citep{doph} modified at the
Bologna  Observatory (by  P.  Montegriffo) to  read  images in  double
precision  format.  A  final  catalogue listing  the instrumental  V,I
magnitudes  for all  the  stars in  each  field has  been obtained  by
cross-correlating the V and  I catalogues. Only the sources classified
as  stars  by  the  code  have been  retained.  The  spurious  sources
erroneously  fitted  by  DoPhot  (as  cosmic  rays,  bright  background
galaxies etc.) have been removed by hand from the catalogues.
In order  to obtain a CMD as  clean as possible, all  the sources with
photometric errors (in V or in I) larger than two times the mean error
at their magnitude were removed from the catalogue, as well as all the
sources with  photometric error  larger than 0.2  mag in at  least one
passband. 
 
The two catalogs were then  matched together: stars in the overlapping
area were  used to  transform the magnitudes  and coordinates  of each
field to a common system. In  particular both the V and I instrumental
magnitudes  and coordinates  for stars  in the  NE Field  catalog were
reported to the  FC system.  In the homogeneous  coordinate system the
cluster center is at $(X_{px},Y_{px}) \simeq (860,920)$.
 
Finally, to minimize the effects of crowding on the definition of the main
branches of the CMD we include in the final catalogue: (a) all the
stars from the FC catalogue with  $V<19.5$ and, (b) all the stars from
the Field NE catalogue more distant than 280 px ($r > 1'.28$) from the
cluster center.
 
The absolute calibration has  been obtained from repeated observations
of \citet{land} standard fields, including all the stars listed in
the extended  catalogue of calibrators provided  by \citet{stet}.  The
details  of the  process  as well as the (successful) tests on the accuracy 
of the calibration will  be described  in  a forthcoming  paper
devoted to  describe the results  of the main  observational program
whose the observing nights  were assigned \citep{mbdra}. The derived
color equations:

$$V = v-0.097(v-i)+26.196$$

$$I = i+0.048(v-i)+25.849$$

(where V,I are calibrated magnitudes and v,i are instrumental magnitudes
reported to 1 s of exposure time and to infinite aperture) provides a zero-point
accuracy of $\pm 0.02$ mag and no detectable residual color effect. 

The final calibrated catalogue is presented in Table 1 where, for each
star, the  following information is listed:  the identification number
(column  1), the  V  magnitude and  error  (columns 2  and  3), the  I
magnitude and error (columns 4 and  5), and the X and Y coordinates in
pixels (columns  6 and 7). Column  8 reports a flag  indicating the RR
Lyrae variables identified in the catalogue of \citet{ba45}, according
to  his  nomenclature. The  stars  with  identification  number $ID  <
10,000$ are from the FC sample, those with $ID\ge 10,000$ are from the
Field NE catalog.

\section{The properties of NGC~5634}

\subsection{The overall CMD morphology }

The final  $(V,V-I)$ CMD  of NGC~5634 is  presented in  Fig.~1.  Stars
with $V <  19.5$ (lying above the dotted line in  Fig.~1) are from the
FC  catalogue, while  those with  $V  > 19.5$  are from  the Field  NE
catalogue.  The main characteristics of the CMD are the following:

(1) The RGB  is well defined  also in the  upper part and it  is quite
 steep  suggesting a  low metal  content for  the cluster  stars.  The
 brightest star in the region of  the giant tip is at $V\sim 14.5$ and
 $(V-I)\sim1.6$), which  is about  3.4 mag.  above  the {\it  Zero Age
 Horizontal Branch} (ZAHB) level  (see \S3.2).  A few Asymptotic Giant
 Branch (AGB) stars are also visible on the blue side of the RGB, from
 $V \simeq 16.0$ to $V \simeq 17.0$.
 
(2) The  HB  is  populated  essentially   on  the  blue  side  of  the
 instability strip, at $(V-I)\le 0.3$. However the cluster is known to
 contain also a few RR Lyrae variables \citep{ba45}.

(3) Despite the not excellent quality of the photometry for $V > 20.5$
 the  Sub  Giant  Branch  (SGB)  and  the Turn  Off  (TO)  region  are
 sufficiently  defined and  the TO  point can  be reliably  located at
 $V_{TO}\simeq 21.2$.
  
(4) The (modest)  effects of  foreground contamination by  field stars
 are mostly  evident around $V-I \simeq  0.4$, for $V>  19.5$, and for
 $V-I \ge 1.2$ at faint magnitudes.

\begin{figure*}
\figurenum{1}
\centerline{\psfig{figure=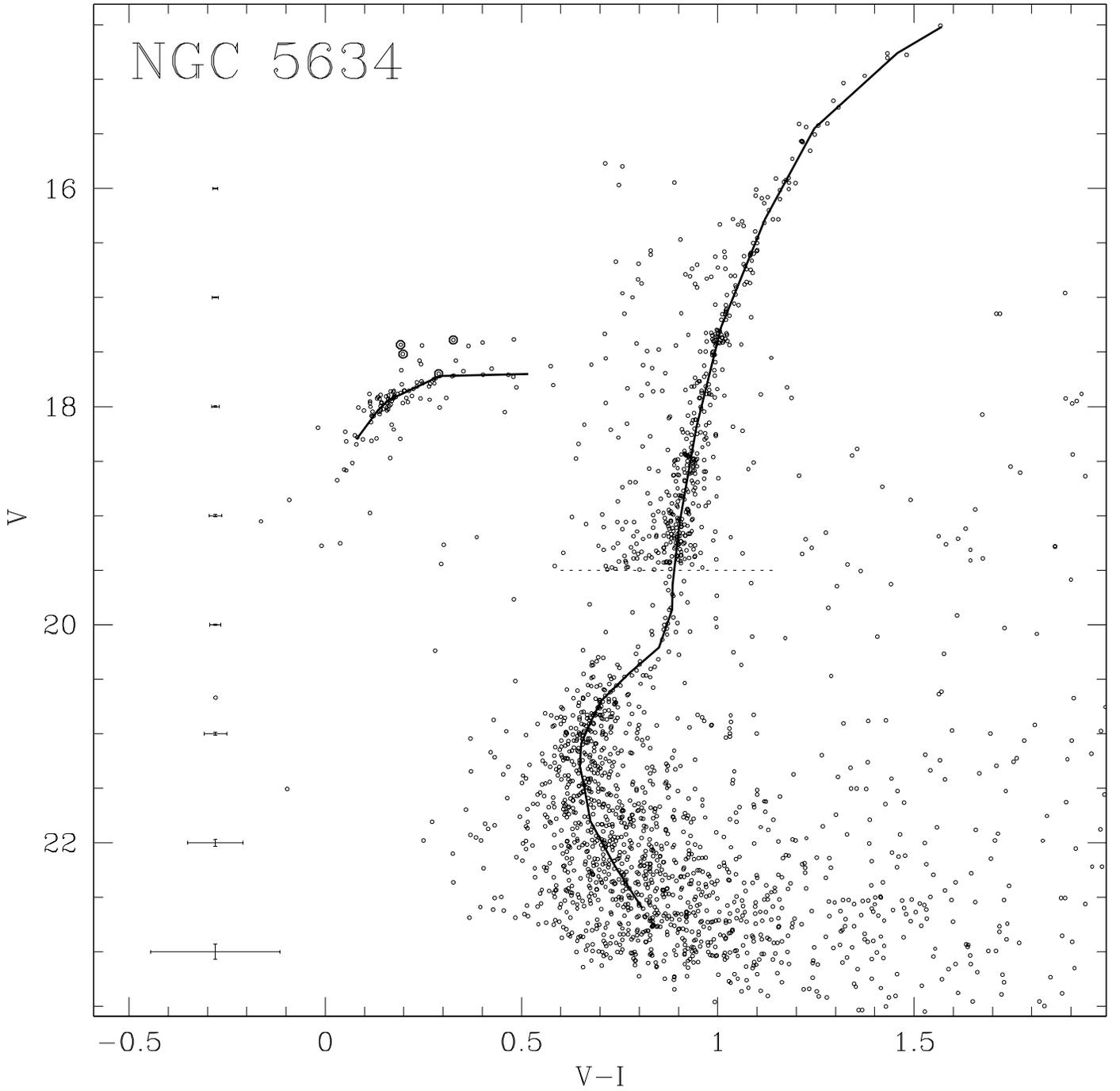}}
\caption{The  final CMD  of NGC~5634,  including 2065  selected stars.
The four RR Lyrae variables  identified in Plate 1 of \citet{ba45} and
included in our catalog are marked with large open circles. The dashed
horizontal line $V = 19.5$  separates the two observed samples (FC and
Field NE  - see text). Note that  only stars measured in  the Field NE
are  considered for  $V >  19.5$.   The mean  ridge line  of the  main
branches  are  over-plotted on  the  CMD.   The  photometric errors  at
different magnitude levels are marked on the left of the diagram.  }
\end{figure*}

We have identified and included in the final catalog four of the seven
RR Lyrae  stars discovered by \citet{ba45},  i.e. variable n.  1, 2, 4
and 6 (hereafter  V1, V2, etc.), that are marked  in Fig.~1 with large
empty circles.   The identification  of V1, V4  and V6 is  safe, while
some  uncertainty  remains for  the  cross-identification of  variable
V2.  V3 was identified  but it  has been  excluded from  our catalogue
because  of  very uncertain  photometry  due  to  blending with  other
sources.  Variable  V7 is  blended in  a group of  stars that  was too
crowded to  be resolved in our  images; V5 was probably  observed at a
phase near the minimum, since it appears  to be very faint in our 30 s
exposures with respect to the other  RR Lyrae, and was not included in
the  final catalogue  because it  did  not pass  the severe  selection
criteria adopted. It  may be interesting to recall that  V1 and V2 are
classified as {\em  ab} type and V4  and V6 as {\em c}  type RR Lyrae,
according  to \citet{lisa}.  According to  their  results \citet{lisa}
classified the  cluster as Oosterhoff  II type, typical of  very metal
deficient globular cluster.

The mean ridge lines of the  main branches in the CMD for NGC5634 have
been obtained by dividing  each sequence in appropriate magnitude (for
the  MS and  the  RGB)  or color  (for  the HB)  bins  and applying  a
$2-\sigma$  clipping algorithm  to each  bin, accordingly  to the
procedure adopted in \citet{mb01}.  The  adopted ridge lines are listed in
Table 2 and plotted (as a heavy solid line) in Fig.~1.
  
\subsection{HB level}

The determination  of the  HB level  in clusters that  have a  blue HB
morphology   is  a   difficult  task.   Moreover,  as   emphasized  in
\citep[][hereafter  F99]{fer99}, some confusion  in the  definition of
the HB level  does exist in the literature since  some papers refer to
mean  level  of  the  HB  or  the  mean  magnitude  of  the  RR  Lyrae
($<V_{RR}>$), other  to the ZAHB  level ($V_{ZAHB}$).  We  recall that
the $V_{ZAHB}$ is  the only value which can  be directly compared with
the theoretical predictions.
    
In the  following we  apply two procedures  in order  to independently
determine the $<V_{RR}>$ and the $V_{ZAHB}$ for NGC5634.

\subsubsection{The $<V_{RR}>$ level}
 
We  adopt here the same procedure followed
by  Montegriffo et  al.   (1998, hereafter  Mo98)  for Ter~8,  another
cluster with  a similar blue HB  \citep[see, also the  case of NGC~288
in][and  discussion  therein]{mb01}.   Following  this  approach,  the
$<V_{RR}>$ level in a cluster with a blue HB morphology is obtained by
the comparison with the HB of a cluster with a well studied population
of RR Lyrae variables, used as a template.  As in the case of Ter~8 we
choose  M68 \citep{w94}  as the  reference cluster.   In panel  (a) of
Fig.~2  the HB  stars of  M~68 (open  dots) are  shifted ($\Delta  V =
+2.04\pm 0.06$  and $\Delta  V-I =  +0.04$) to match  the HB  stars of
NGC~5634 (full dots).
The thick dashed  line marks the mean level of  the RR Lyrae variables
in M~68  \cite[$V_{RR} = 15.64  \pm 0.01$][]{w94} once shifted  by the
same amount,  thus giving $V_{RR}^{NGC5634}  = 17.68 \pm  0.06$.  This
value is  used in  Section 3.3  to determine a  set of  RGB parameters
which depend on the metal content and the reddening of the cluster.

In  {\it panel (b)}  of Fig.~2,  the result  of a  further consistency
check is  shown: the HB of  Ter~8 (open squares) has  been shifted (by
$\Delta V = -0.28 \pm 0.08$ and  $\Delta V-I = -0.06$) to match the HB
of NGC5634  (full dots).  As can be  seen, the match is  very good: by
appling the shift in magnitude  obtained above ($\Delta V = -0.28$) to
the  mean RR  Lyrae level  of NGC5634  ($V_{RR}^{NGC5634} =  17.68 \pm
0.06$) we  obtain, for Ter~8, $V_{RR}^{Ter8}=17.96\pm  0.06$, a figure
which   is   fully  consistent   with   the   value   found  by   Mo98
($V_{RR}^{Ter8}=17.95\pm 0.05$). This test demonstrates the good degree of 
repeatability of the whole ``HB matching'' technique adopted here.

\begin{figure*}
\figurenum{2}
\centerline{\psfig{figure=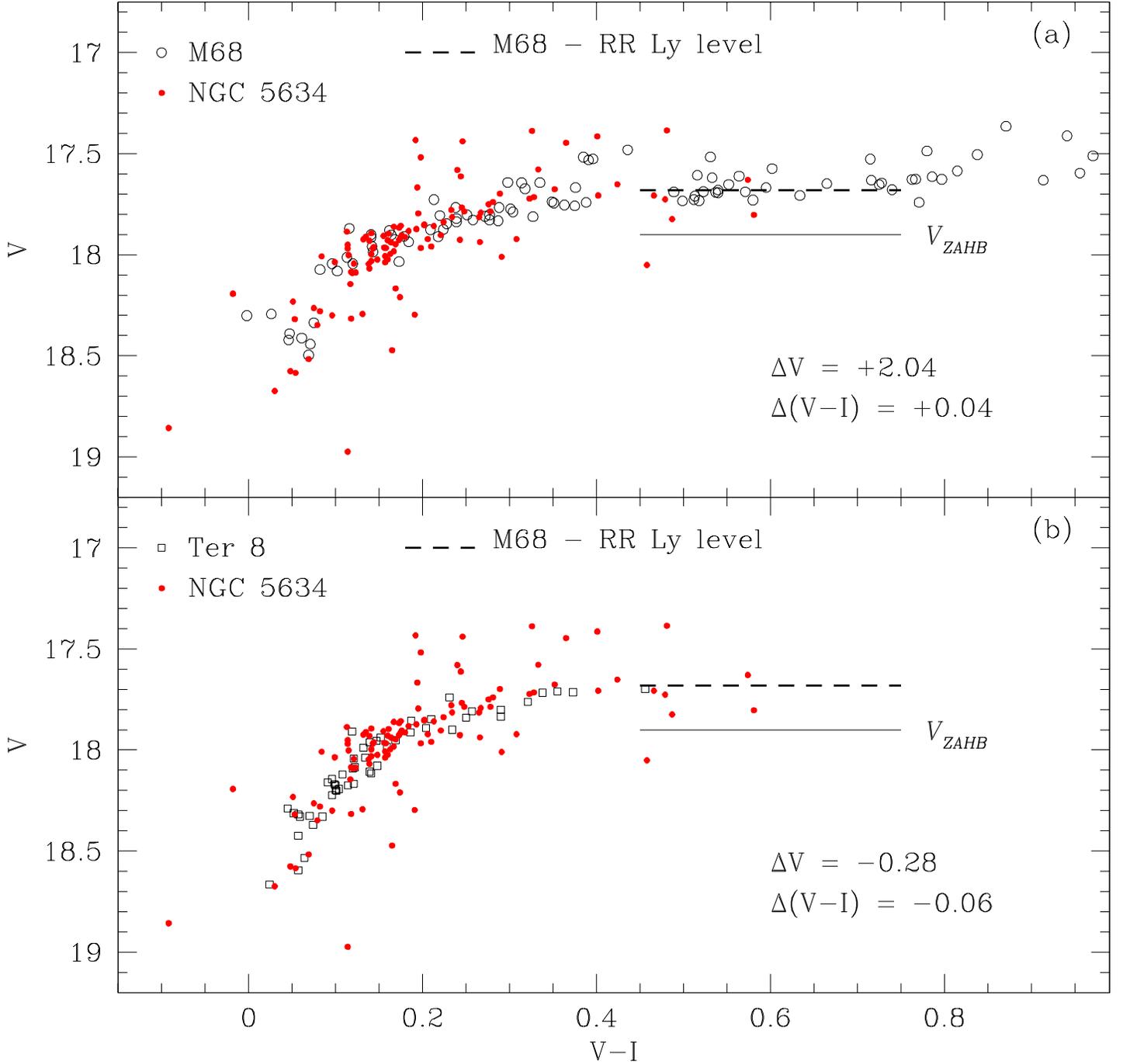}}
\caption{ The HB level in NGC5634.   {\it Panel (a)}: HB stars of M~68
\cite[open circles;  data from][]{w94} have been shifted  to match the
HB of NGC~5634 (solid circles).  The shift in magnitude and colors are
indicated.   The  dashed line  is  the mean  RR  Lyare  level of  M~68
(estimated by \citet{w94}) after the magnitude shift has been applied.
The  solid  line is  the  ZAHB  level  obtained following  the  method
described  in  F99.   {\it  Panel  (b)}:  The  same  for  Ter~8  (open
squares).  The two levels  ($<V_{RR}>, V_{ZAHB}$)  are reported  as in
{\it panel (a)}.  }
\end{figure*}

\subsubsection{The $<V_{ZAHB}>$ level}

We  also determine the level of the Zero
Age Horizontal  Branch (ZAHB)  at $Log T_e  =3.85$, following  the the
semi-empirical approach  adopted by F99.   The method is based  on the
comparison of  synthetic and observed HB.  The  observed HB morphology
is  reproduced by means  of synthetic  HB diagrams  (see F99  for more
details). Once  the observed  HB distribution is  properly reproduced,
the synthetic HB is shifted in order to match the observed one and the
ZAHB level at  $Log T_e =3.85$ is read from  the shifted sythethic HB.
The procedure applied to NGC~5634 yelds $V_{ZAHB} = 17.90 \pm 0.10$.

As above, an additional consistence check can be done with the cluster
Ter8. In  fact, by coupling the  value of the ZAHB  obtained above for
NGC5634  and   the  value   listed  in  Table   2  by  F99   for  Ter8
($V_{ZAHB}^{Ter8}   =    18.15   \pm   0.10$)    we   obtain   $\Delta
V_{ZAHB}(NGC5634-Ter8)=  -0.25\pm 0.10$,  in excellent  agreement with
the difference in $<V_{RR}>$ found above.

The difference between $V_{ZAHB}$ and $<V_{RR}>$ levels for NGC5634 is
$\delta  V(ZAHB-RR)\sim  0.22$   mag,  confirming  that  a  significant
difference  between  the  two  levels  does indeed  exist  in  blue-HB
morphology clusters (see F99).   Note also that $\delta V(ZAHB-RR)$ is
larger  that what  is  expected by  using  relation (2)  in F99.  This
evidence confirms once more that such relations should be used to have
only a  first-guess estimate, since the actual  difference between the
two  levels  strictly  depends  on   the  HB  morphology,  and  it  is
particularly large for clusters with essentially blue HB morphology.

The derived  $V_{ZAHB}$ also  suggests that the  reddest part  ($V-I >
0.3$)  of the BHB,  and possibly  the bluest  part of  the instability
strip, may  be populated by evolved stars  significantly brighter than
the ZAHB,  similar to the  case of NGC~288  \citep[see][for discussion
and references]{cat,don,frank}.

\subsection{Metallicity and reddening}

From  integrated   colors  (in  particular   from  the  reddening-free
parameter Q39) \citet{Z80} derived  a reddening $E(B-V)=0.04 \pm 0.03$
while  \citet{rhs}  report  $E(B-V)=0.06  \pm 0.03$.   With  the  same
parameter   \citet{ZW84}  estimated   $[Fe/H]_{Q39}=-1.87$,   while  a
slightly  higher  value  is  reported  from  $\Delta  S$  measures  by
\citet{sp82},  \citep[see  also][]{pila}.   The  final  average  value
reported  by \citet{ZW84} is $[Fe/H]=-1.82  \pm 0.13$, which takes also into 
account the fact that the scale adopted by \citet{sp82} gives metallicity
values sistematically higher  by  $\sim   0.2$  dex  with  respect  
to   those by \citet{ZW84}. The original $[Fe/H]_{Q39}=-1.87$  
is probably the most consistent figure to compare with.

We  use the photometry presented here in order to derive an independent
photometric estimate of the cluster metallicity from the position and the
morphology of the giant branch. In the following we adopt E(V-I)=1.34E(B-V),
according to \citet{dean}. 

\citet{smr} introduced a method  (SMR method) to derive simultaneously
the metallicity and reddening of a  globular cluster by using a set of
relations   tieing   $[Fe/H]$   with   two  observables:   the   color
[$(V-I)_{0,g}$]  and the slope  [$\Delta V_{1.2}$]  of the  RGB. Since
$(V-I)_{0,g}$ depends  on reddening  while $\Delta V_{1.2}$  does not,
the procedure provides  also the value of the  reddening for which the
consistency between the two  independent relations is reached. The SMR
method  has also  been  recently calibrated  in the  \cite[][hereafter
CG97]{cg97} metallicity scale by \citet{cb98}.

A clear and  robust solution is found here  for both the calibrations,
giving  $[Fe/H]_{ZW84} = -1.98  \pm 0.10$  , and  $E(B-V) =  0.060 \pm
0.010$ for the \citet{smr} calibration, and $[Fe/H]_{CG97} = -1.94 \pm
0.10$ , and $E(B-V) = 0.063  \pm 0.010$ for the \citet{cb98} one.  The
reddening estimate is in good agreement with the values reported in the
literature as  well as with  the estimate obtained from  the IRAS-COBE
maps  \cite[][$E(B-V) = 0.056  \pm 0.005$]{sf}.   In the  following we
adopt $E(B-V) =  0.06 \pm 0.01$.  The derived  metallicity in the ZW84
scale  is  slightly  lower  than  what reported  by  ZW84,  but  still
compatible within the errors.

In  order to  be  fully  consistent with  the  assumptions and  scales
adopted in F99,  we also determine the global  metallicity ([M/H]) (as
defined by eq 1 in  F99).  By assuming $[Fe/H]=[Fe/H]_{CG97} = -1.94$,
the global metallicity turns out to be $[M/H]=-1.74$.

\subsection{The RGB bump}

The RGB-Bump is an evolutionary  feature occurring along the RGB which
flags  the  point  where  the  H-burning shell  crosses  the  chemical
discontinuity  left  by  the  maximum penetration  of  the  convective
envelope.  From the  observational point of view it  was only recently
identified     in     a     significant     number     of     clusters
\citep[][F99]{ffp90,manu99}.

The change in the slope of the cumulative luminosity function (LF) and
the excess  of star counts in the  differential LF of the  RGB are the
main  tools  to identify  the  RGB-Bump.  In particular  \citet{ffp90}
suggest  that the  change in  the slope  of the  cumulative LF  is the
safest indicator of the RGB-Bump.

In Fig.~3  the RGB LF  for NGC5634 in  the bump region is  shown.  The
change in the slope of the  cumulative LF is shown in {\it panel (b)}.
As can be seen it corresponds  to the obvious peak in the differential
LF shown in {\it panel (a)}.  Thus, the RGB-Bump is clearly identified
at $V_{RGB}^{bump} =  17.38 \pm 0.05$.  A clear clump  of RGB stars at
that magnitude is clearly visible also in the CMD shown in Fig.~1.

Our determination of the RGB-Bump magnitude is significantly different
from the  value listed by  \citet{manu99}, who presented  the RGB-Bump
magnitudes for 28 GGCs, from the
HST {\em  snapshot survey}. For NGC5634 they  listed $V_{RGB}^{bump} =
17.77  \pm 0.03$.   The  large discrepancy  ($\Delta  V\sim 0.4$  mag)
suggests  a  serious problem  in  the zero-point  in  one  of the  two
photometries.   However, since our  calibration was  carefully checked
against other  well-calibrated fields  observed during the  same night
(see \S2.3), we regard our measure as the most reliable one.  Moreover
a new re-calibration  of the HST data-set seems  to suggest a RGB-Bump
magnitude for NGC 5634 significantly brighter than that published
in \citet{manu99} (Piotto 2002 - private communication).

\begin{figure*}
\figurenum{3}
\centerline{\psfig{figure=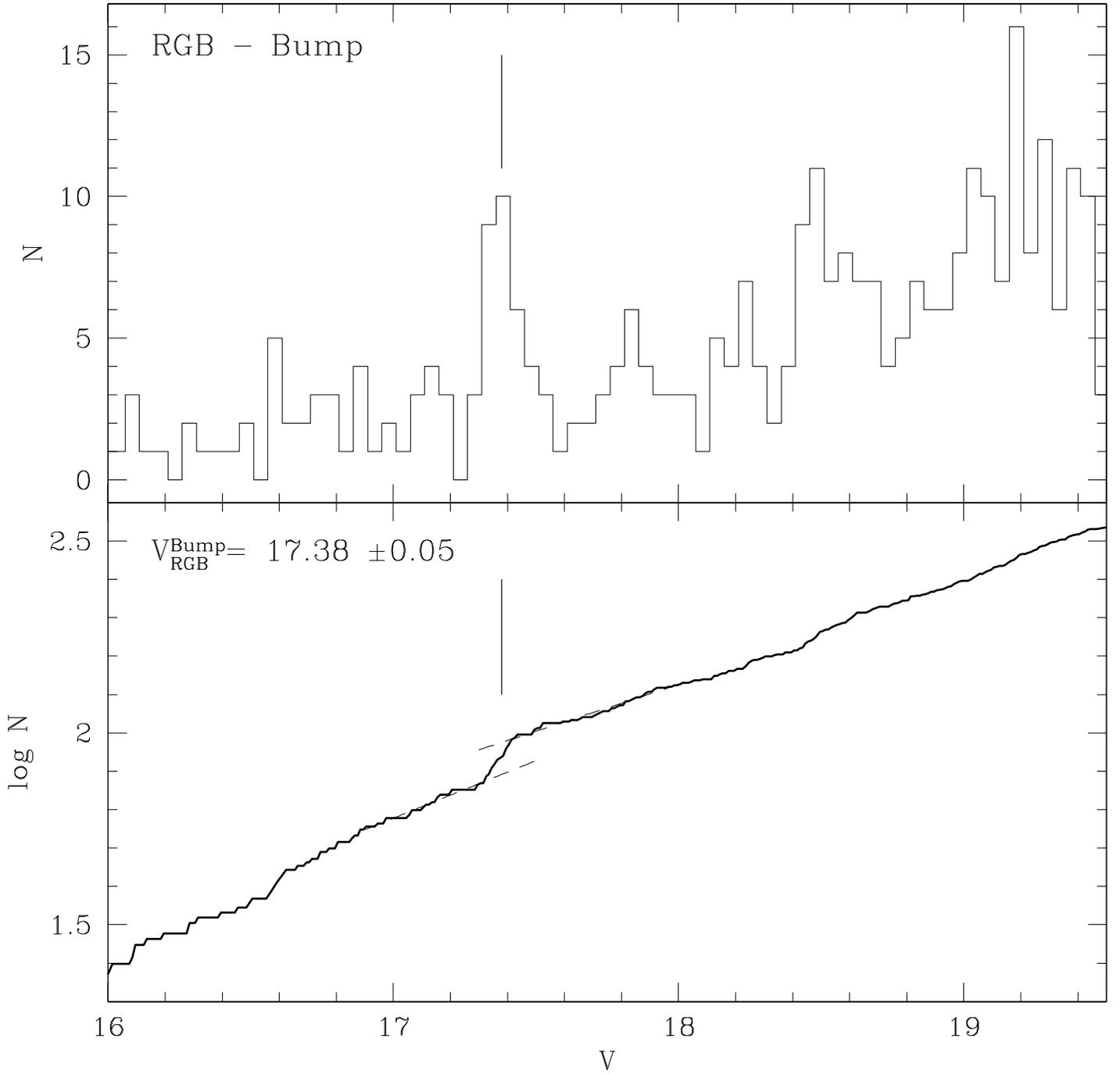}}
\caption{Panel (a): the differential luminosity function of the RGB of
NGC~5634. The peak  indicated by the vertical line  corresponds to the
RGB bump.  The  same feature is identified in Panel  (b) by the sudden
change in slope of the logarithmic cumulative luminosity function. }
\end{figure*}

The magnitude  of the RGB-bump with  respect to the  HB level ($\Delta
V_{HB}^{bump}$) has been  used by \citet{ffp90} (and later  by F99 and
\citet{manu99})  to  compare  the  observations with  the  theoretical
predictions.   The  $\Delta   V_{HB}^{bump}$  parameter  is  mainly  a
function of the  metallicity and it has only a  mild dependence on age
\citep[see][F99]{ffp90}.  Adopting the  ZAHB level obtained in Section
3.2,   $V_{ZAHB}=17.90\pm0.10$  we   obtain  for   NGC   5634  $\Delta
V_{HB}^{bump}=-0.52  \pm 0.08$.   Fig.~4 shows  the comparison  of the
value obtained for NGC~5634 with 47 GGCs from the data-base by F99, in
the  $[Fe/H]$  vs.  $\Delta  V_{HB}^{bump}$ plane.   The  position  of
NGC~5634 in  this plot is in  excellent agreement with  the mean locus
defined by the other  Galactic globulars, independently of the assumed
metallicity scale (see panels a and b of Fig.~4).

\begin{figure*}
\figurenum{4}
\centerline{\psfig{figure=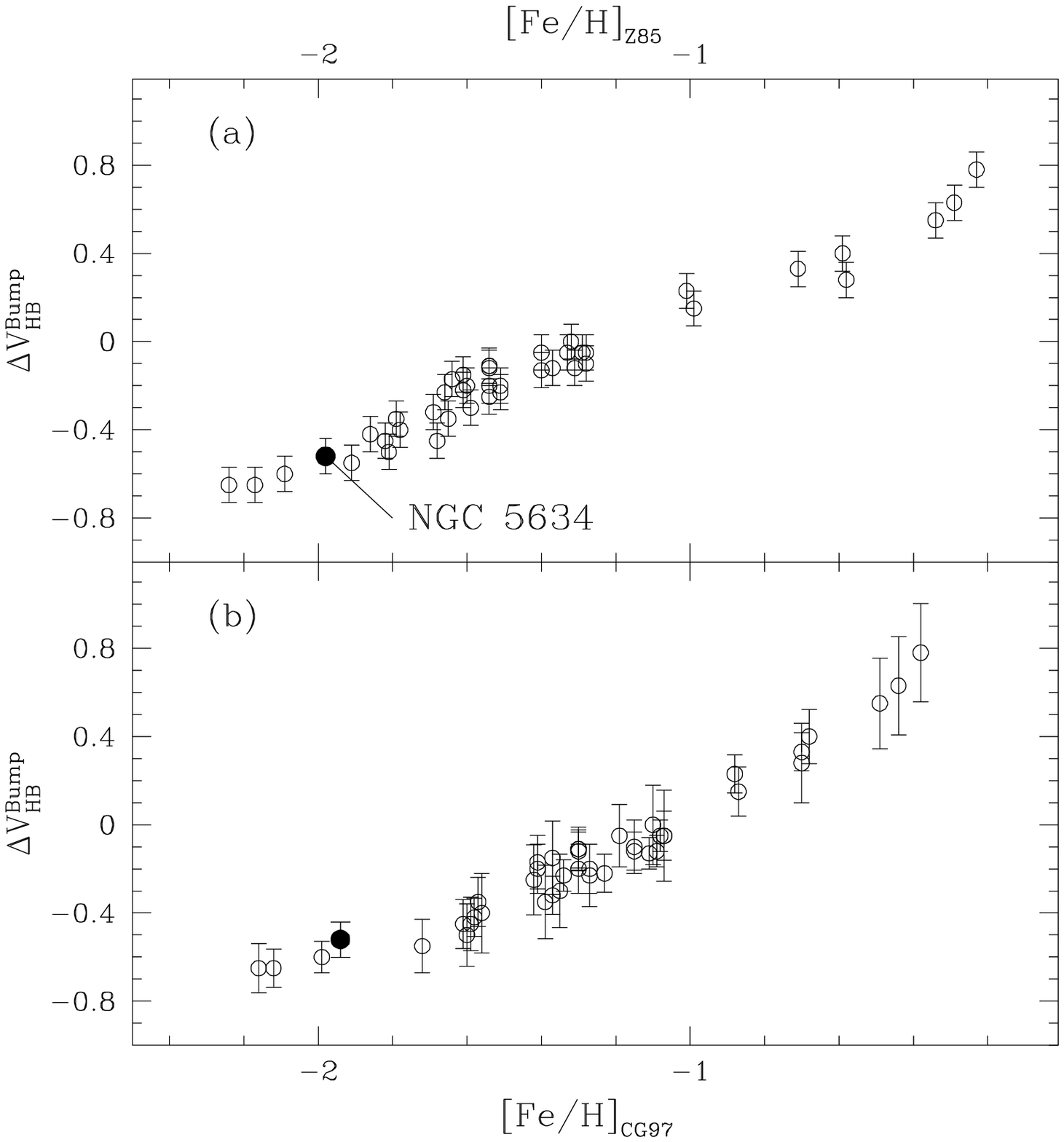}}
\caption{The difference in magnitude between the RGB-Bump and the ZAHB
level ($\Delta  V_{HB}^{Bump}$), as a  function of the  metallicity in
the Zinn scale ($[Fe/H]_{Z85}$) and in the CG97 scale, {\it Panel (a)}
and {\it (b)},  respectively. The filled circle is  NGC5634, the empty
circles are the clusters from Table 5 of F99.}
\end{figure*}

\section{Distance and age}

In deriving the  distance to NGC~5634 we use  (following F99) the ZAHB
level  as  primary  standard  candle:  by using  the  F99  calibrating
relation  (their Eq.~5,  $M_V^{ZAHB}=0.23[M/H]+0.94$)  and the  global
metallicity obtained in \S3.3 we derive $M_V^{ZAHB}=0.54$. Coupling
this value with the  observed ZAHB level ($V_{ZAHB}=17.90\pm 0.1$, see
\S3.2) and the reddening ($E(B-V)=0.06\pm 0.01$, see \S3.3), we obtain
for NGC~5634 a {\it true} distance modulus $(m-M)_0 = 17.17 \pm 0.12$.

We can check this result using the RGB  bump as a standard candle, with
the empirically-calibrated relations between $M_V^{bump}$ and
metallicity presented by F99 (their Eq.~6.7 and 6.8). 
In both cases we  obtain $(m-M)_0 = 17.22 \pm 0.34$, in excellent agreement
with the above estimate.

Finally, we determined the  distance modulus differentially
with  respect to  M~68.  By  adopting the  shift found  in  \S3.1, the
distance  modulus  and reddening  of  M~68  given  by F99,  we  obtain
$(m-M)_0 =  17.09 \pm 0.12$, also in agreement (within  the uncertainties)
with our final estimate.

\subsection{The Relative Age of NGC~5634}

A number  of observational evidences  (such as the low  metal content,
the blue HB morphology and the  position of the RGB bump) suggest that
NGC~5634 is  an old cluster.   However a quantitative estimate  can be
obtained  only from  the direct  measure of  the key  age-diagnostic of
stellar populations: the MS Turn Off (TO) point.

As  is well known  two main  methods are  currently used  to determine
relative  ages for  GGCs:  the  {\it vertical  method},  based on  the
measure  of  the  parameter  $\Delta V^{HB}_{TO}$  \citep{bcf},  which
measure the luminosity  of the TO point with respect  to the HB level;
and  the {\it  horizontal method}  based  on the  difference in  color
between the TO and a point along  the RGB 2.5 mag brigther than the TO
point \citep{svb96}.
  
As  has already been  emphasized elsewhere,  the horizontal  method is
particularly  sensitive  to color  uncertanties:  even small  ($\simeq
0.02$ mag)  errors generate  large ($\sim 2$  Gyr) differences  in the
derived  relative  ages.  Since  our  photometry  is not  sufficiently
accurate for $V>20.5$, to derive a precise measure of the TO color, in
the  following we rely  only on  the vertical  method to  estimate the
relative  age of  NGC5634  and we  will use  the  TO color  only as  a
consistency check.

From the ridge line listed  in Tab. 2 we obtain $V_{TO}=21.22\pm 0.15$
and $(V-I)_{TO}=0.647\pm 0.03$.  Adopting  the value of the ZAHB level
($V_{ZAHB}$)  derived above, we  get ($\Delta  V^{HB}_{TO} =  3.32 \pm
0.16$). This value is in good agreement with the bulk of the other old
and  metal  poor  Galactic  globulars, as  homogeneously  measured  by
\citet{alf} \footnote{To  preserve the homogeneity with  the age scale
of \citet{alf},  the vertical parameter  are computed with  respect to
the ZAHB level.}.  The {\em  horizontal} age parameter turns out to be
$\delta  (V-I)_{@2.5} =  0.272 \pm  0.04$, a  value which  is  also in
agreement - within the errors - with other old and metal poor Galactic
globulars.

\citet{alf}, for  example, report for  M68 $\Delta V^{HB}_{TO}  = 3.30
\pm 0.12$  and $\delta (V-I)_{@2.5}  = 0.306 \pm 0.010$.  These values
are also  consistent with what found  in Ter~8. In fact, using the
results  by Mo98 and  F99, we  obtain $\Delta  V^{HB}_{TO} =  3.40 \pm
0.14$ and  $\delta (V-I)_{@2.5}  = 0.301 \pm  0.04$ for  Ter~8.  These
values are very similar to those derived for NGC~5634, suggesting that
the  cluster is as  old as  M~68 and  Ter~8 (within  the observational
uncertainties of $\sim \pm 2 Gyr$).

\section{NGC~5634: a cluster in the Sgr Stream}

Since its discovery  \citep{s1} the Sgr dSph has  been suspected to be
the damaged  relic of  a fomerly larger  stellar system that was
harrassed by the tidal field of our Galaxy. This view has now received
many   observational   confirmations   \cite[see,   for   example][and
references therein]{sdss}.   The stars lost by  a disrupting satellite
are expected  to be distributed along  its orbital path, in  a sort of
stellar {\em stream} more or less spread around the orbit depending on
the degree of  flattening of the Galactic dark halo  and on the effect
of   the  gravitational   potential   of  the   disc  \cite[see,   for
instance,][]{katy1,katy2,orb1}. In the  last few years, many different
teams have found clumps and/or  density structures in the Galactic halo
that  are  probably associated  with  the  stream  of the  Sgr  galaxy
\citep{vivas,dohm,yanni,ivez,orb2,rod,maj,mat,alard96}   and   in  two
cases   a   clear   and   direct   association   has   been   obtained
\citep{david,sdss} by  comparing the stellar population  of the stream
with that  of the Sgr  galaxy.  From these  studies it seems  that the
stars  lost  by  the  Sgr  galaxy  are distributed  in  a  giant  halo
substructure, a  low density band  probably encircling the  whole sky,
approximately following the past orbital path of Sgr \citep{rod,sdss}.

Since  the main  body of  the Sgr  dSph still  encloses  four globular
clusters  (M~54, Ter~8,  Arp~2,  and Ter~7)  that  ultimately will  be
incorporated into the globular cluster system of the Milky Way, it may
be presumed  that also some previous  Sgr globular may  have been lost
along the  Sgr stream. A strong  case for this  occurrence has already
been  made  for  Pal~12 on  the  basis  of  its present  position  and
three-dimensional  velocity  \citep{dana}.   Here  we  show  that  the
galactocentric  position  and  the  radial velocity  of  NGC~5634  are
compatible with  the hypothesis that  also this cluster  is associated
with the Sgr stream and it was a former member of the Sgr galaxy.

In  Fig.~5 the  Galactic globular  clusters of  the outer  halo (i.e.,
those with  $R_{GC}\ge 10 $ kpc)  are reported as open  circles in the
planes $R_{GC}$  vs. $V_r$ (panel  a), X vs.  Z, X vs.  Y and Y  vs. Z
(panels b, c, and d, respectively), where $V_r$ is the radial velocity
in km/s and X,Y,Z are galactocentric coordinates in kpc.
 
The  filled circles are  the Sgr  globulars and  the filled  square is
Pal~12. The  large encircled dot is  NGC~5634.  In the  same planes we
have also  reported the orbital path  followed by the Sgr  dSph in the
last  Gyr (from  1 Gyr  ago to  the present),  according to  the orbit
computed  by \citet{orb1},  based  on the  proper  motion measured  by
\citet{s2,orb2}.  The small arrow at one end of the orbital path marks
the position (and velocity) of the Sgr galaxy 1 Gyr ago.

\begin{figure*}
\figurenum{5}
\centerline{\psfig{figure=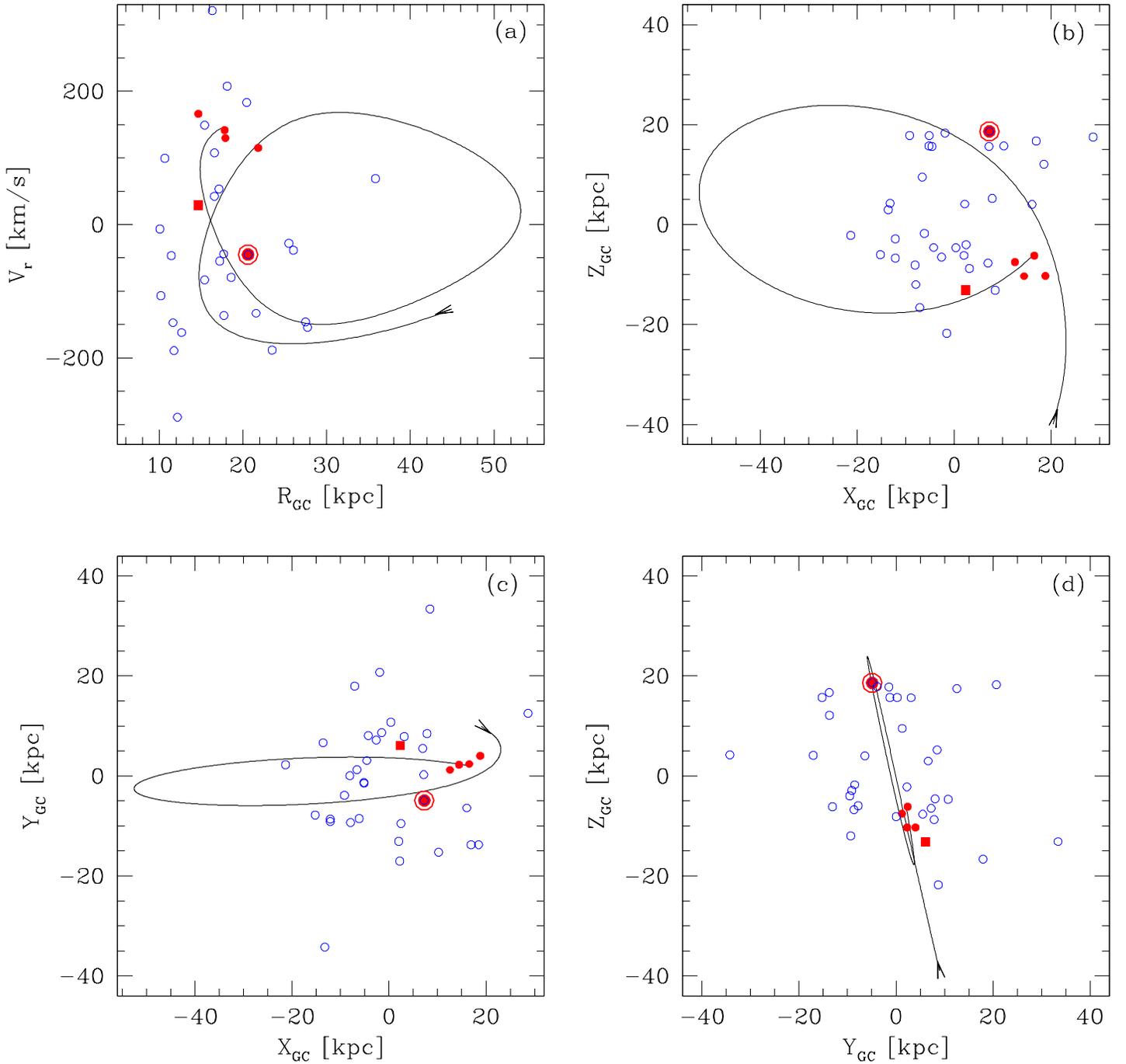}}
\caption{The  outer   Galactic  globular  clusters   (with  $R_{GC}\ge
10$~kpc) are plotted  (open circles) in the $R_{GC}$  vs.  $V_r$ plane
({\it panel (a)})  and in three different projections  in the space of
galactocentric coordinates  (X vs.  Z,  X vs. Y  and Y vs.  Z  in {\it
panels  (b), (c)},  and {\it  (d)}, respectively).   All  the globular
cluster  data are  from  the \citet{harris}  catalogue.  The  globular
clusters  presently associated  with the  main body  of the  Sgr dSph,
i.e. M~54, Ter~8, Arp~2 and Ter~7 are plotted as small filled circles.
The position of NGC~5634 is marked as a large filled circle.  The full
square is  the globular  cluster Pal~12 that  has been indicated  as a
former member of the Sgr  galaxy by \citet{dana}.  The continuous line
is the  orbital path  followed by  the Sgr dSph  during the  last Gyr,
according to the  orbit computed by \citet{orb1}. The  arrow marks the
point corresponding (in the simulation)  to 1 Gyr ago, while the other
end of the path corresponds to the present.}
\end{figure*}

NGC~5634, as well  as Pal~12, lie near the predicted  orbit in all the
planes displayed in  Figure~5.  This means that {\em  its position and
radial  velocity are  fully compatible  with the  hypothesis  that the
cluster is following the same orbit  of the Sgr dSph}, i.e. it belongs
to  the  Sgr Stream  \footnote{  We  have  found evidence  that  other
clusters share these properties  in the (X,Y,Z,$V_r$) space.  The case
is discussed in detail in a dedicated paper, \citep{mbstream}.}.  Note
that in all the considered  planes NGC~5634 and Pal~12 have a distance
from the orbit that is lower than the spread of confirmed Sgr clusters
around the orbit itself.

While  an ultimate  confirmation  - at  least  for NGC~5634  - can  be
achieved only with accurate proper motion and a more secure assessment
of  the  Sgr orbit\footnote{Another possible observational test would be the
search for Sgr stream stars in a wide field around the cluster, lying at the
same distance from us as the cluster itself \citep[see, e.g.][]{david12}.},  
the  evidence is  very  suggestive  and the  idea
deserves  serious consideration.  It  would mean  that NGC~5634  was a
former member of the globular cluster  system of the Sgr dSph and that
it formed in  that stellar system. In the following  we will take this
association as a  working hypothesis and we will  shortly discuss some
interesting clues:

\begin{itemize}

\item We remark the already  noted similarity of NGC~5634 with the Sgr
globular  Ter~8.  Considering  the stellar  content, the  two clusters
appear as ``identical twins''. They  share the same metal content, the
same HB morphology and the same age.

\item The stellar  population of NGC~5634 appears more  similar to the
population found in the Sgr  Stream \citep{sdss} than that observed in
main body of the Sgr galaxy.  \citet{sdss} found that the stars in the
Sgr  stream  show a  BHB  population (similar  to  that  of Ter~8  and
NGC~5634)  much more  pronounced than  what  is found  in the  central
regions of the galaxy.

\item \citet{sgr1} found evidences of the  presence of an old and very metal
poor  population (similar  to that  of Ter~8)  in the  Sgr dSph. The
existence of this population  has been confirmed by \citet{cse} and
\citet{carla}.
Furthermore, many independent studies
\citep{sl95,sgr1,sgr2,ls00,alard01,carla}  have found  evidence  of a
metallicity (and population)  gradient in the Sgr dSph,  with the most
metal  poor stars  being more  abundant in  the outer  regions  of the
galaxy.

\end{itemize}

All the above evidence is consistent with a scenario in which the more
external stars of  the Sgr dwarf are preferentially  stripped from the
galaxy and thus, the Sgr stream is populated by a stellar mix in which
the metal  poor stars  are more abundant,  because of  the metallicity
gradient.   NGC~5634 should  have the  same origin  as the  metal poor
population of the  Sgr galaxy, so it is not  surprizing that it shared
the same ``dynamical  destiny'' as most of the metal  poor halo of the
Sgr dSph, and it is now lost in the Sgr stream.

If this interpretation of the  heritage of NGC~5634 is correct, it has
interesting consequences for our  understanding of the Sgr stream. One
important difference  between NGC~5634, and  the previously identified
clusters (and former clusters) of the  Sgr system, is that it lags the
main body of the dwarf galaxy by an exceedingly large distance. Taking
the IL98 orbit,  we find that the lag along  the orbital trajectory is
$151\kpc$.  To separate  tidal debris from the main  body of the dwarf
galaxy by  such a  large distance requires  a considerable  time.  The
internal velocity dispersion within the bound core of the Sgr dwarf is
$11.4\pm0.7\kms$ \citep{s2},  so if the material  that becomes unbound
acquires a similar velocity dispersion, it would take $\sim 13\Gyr$ to
spread out over  $151\kpc$. A better estimate of  this separation time
can  be made  from numerical  experiments.   In Figure~6  we show  the
end-point structure of the  best-fit simulation in \citet{orb2}, where
we have plotted only those  points which were bound $4\Gyr$ ago (right
panel). The  downstream tail nearly  reaches the position  of NGC~5634
(marked with  an asterisk).   Since particles that  were bound  to the
dwarf galaxy model less than  $4\Gyr$ ago are unable to disperse along
the orbit  to the  position of the  globular cluster,  this simulation
suggests  that NGC~5634  became  unbound more  than  $4\Gyr$ ago.   (A
significantly  longer time, $\sim  6\Gyr$, is  required for  more than
10\% of the stream to drift  beyond the distance of the star cluster).
This  sets a  useful  limit on  the  disruption history  of the  dwarf
galaxy.  It would suggest that  no significant deflection of the orbit
of Sgr occurred  during this time, contary to what  is required by the
scenario  envisaged by \citet{zhao},  who proposed  that Sgr  was shot
into its present orbit, where  it suffers vigorous disruption from the
Galactic tides, following a  close encounter with the LMC $2$--$3\Gyr$
ago.   It  also  places  a   strong  constraint  on  the  proposal  by
\citet{cse}, that Sgr  is a structure that was torn  off from the LMC,
since the time  of that incident has to have  happened before the time
when NGC~5634 became unbound from the Sgr galaxy.

\begin{figure*}
\figurenum{6}
\centerline{\psfig{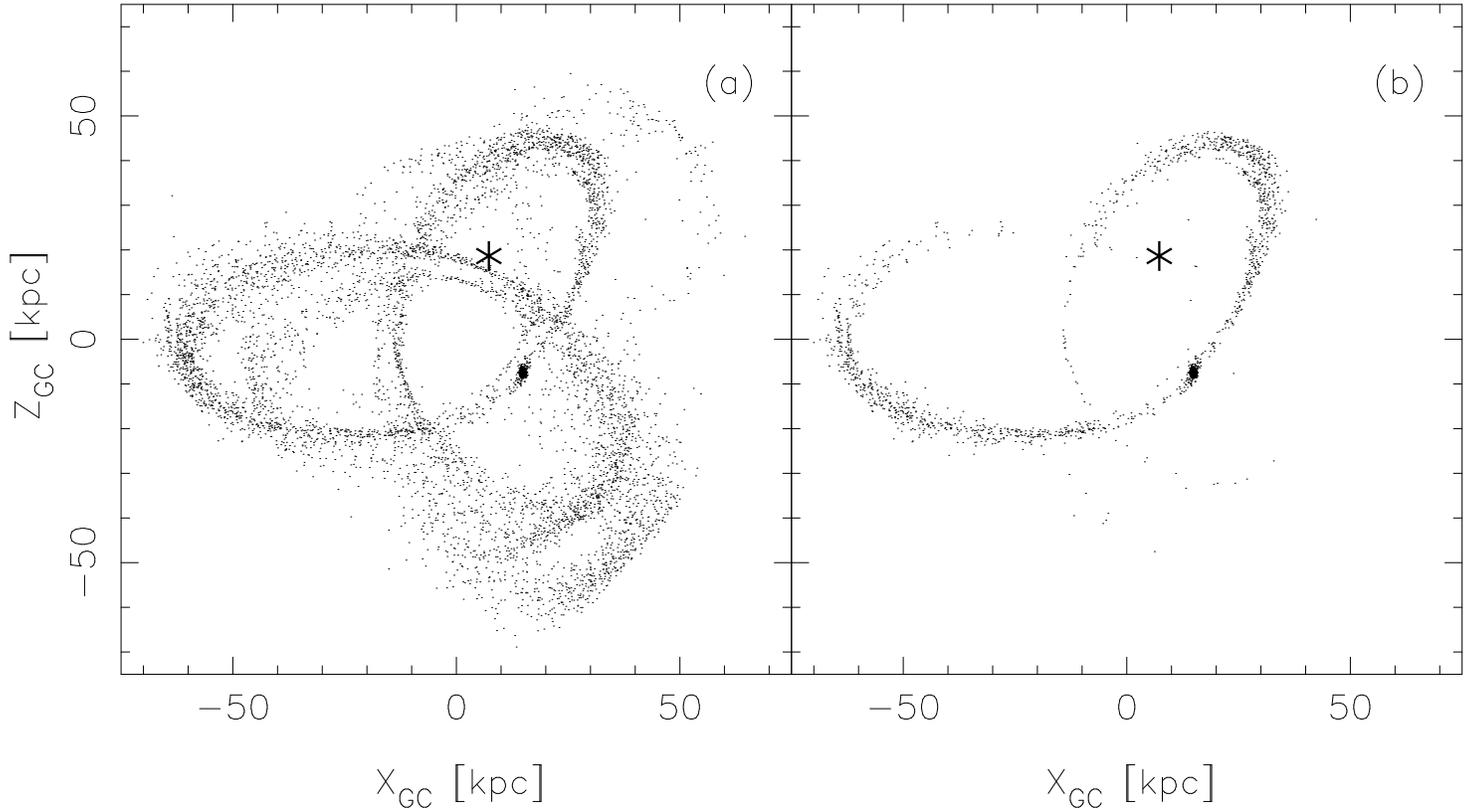}}
\caption{The structure of the tidal  stream of the Sgr dwarf galaxy in
the Galactic $X$-$Z$ plane  (with the Sun located at $X_{GC}=-8\kpc$),
according to  the best-fit simulation of \citet{orb2}  which retains a
bound center  at the  present day. The  simulation corresponds  to the
\citet{orb2} dwarf galaxy model is ``D1'', and halo model ``H2'', in a
Galactic potential  with circular velocity  $v_c=200\kms$ at $50\kpc$.
The left-hand panel displays  all particles, whereas in the right-hand
panel  only those  particles that  were gravitationally  bound  to Sgr
$4\Gyr$ ago are plotted.   Particles that became unbound more recently
do not reach the position of NGC~5634 (indicated by an asterisk), from
which we  deduce that if NGC~5634  was previously a member  of the Sgr
system, it must have parted company with that galaxy more than $4\Gyr$
ago.}
\end{figure*}

\acknowledgments

This research has been  partially supported by  the
{\it Agenzia Spaziale Italiana} (ASI)  and by  the 
{\it Ministero dell'Istruzione,  dell'Universit\`a e della Ricerca Scientifica} 
(MIUR) 
through the grant p. 2001028879, assigned to the project {\em
Origin and Evolution of Stellar Populations in the Galactic Spheroid}.
 We are grateful to the staff of 
the Telescopio Nazionale Galileo (TNG)
for the warm hospitality and the professional assistance.
Part of the data analysis has been performed using 
software developed by P. Montegriffo at the Osservatorio Astronomico
di Bologna. This research has made use of NASA's
Astrophysics Data System Abstract Service.  


\clearpage

\begin{deluxetable}{lccccccc}
\tablecolumns{8}
\tablewidth{0pc}
\tablecaption{Photometry of NGC 5634}
\tablehead{
\colhead{ID} & \colhead{V} &\colhead{$\epsilon_V$}  & \colhead{I} & 
\colhead{$\epsilon_I$} & \colhead{$X_{[px]}$} & \colhead{$Y_{[px]}$} &
\colhead{Var. n.}\\}
\startdata
     1 & 14.505 & 0.002 & 12.938 & 0.007 & 1207.79 &  703.35 & 0\\
     2 & 14.776 & 0.003 & 13.295 & 0.008 &  497.33 &  913.35 & 0\\
     3 & 14.761 & 0.002 & 13.329 & 0.005 & 1060.98 & 1242.67 & 0\\
     4 & 14.802 & 0.003 & 13.369 & 0.006 &  911.93 & 1165.44 & 0\\
     5 & 14.967 & 0.002 & 13.593 & 0.010 &  684.00 &  535.87 & 0\\
     6 & 15.032 & 0.003 & 13.711 & 0.011 &  879.92 &  945.47 & 0\\
     7 & 15.196 & 0.003 & 13.901 & 0.006 &  863.80 &  863.90 & 0\\
     8 & 15.258 & 0.003 & 13.950 & 0.005 &  887.75 &  917.10 & 0\\
     9 & 15.404 & 0.003 & 14.125 & 0.005 &  987.92 &  859.35 & 0\\
\nodata&\nodata&\nodata&\nodata&\nodata&\nodata&\nodata&\nodata\\
\enddata
\tablecomments{A sample of the photometric catalog. The complete catalog is
available in ASCII format in the electronic edition of the paper. The variable
stars are numbered after the nomencalture by \citet{ba45}}
\end{deluxetable}


\begin{deluxetable}{lcclc}
\tablecolumns{4}
\tablewidth{0pc}
\tablecaption{Ridge lines of the RGB,MS and HB.}
\tablehead{
\multicolumn{2}{c}{RGB+MS} &
\multicolumn{1}{c}{} &
\multicolumn{2}{c}{HB} \\
\colhead{V} & \colhead{V-I}&\colhead{}&\colhead{V-I} & \colhead{V} \\}
\startdata
  14.52  &  1.570 & & 0.080  & 18.30\\
  14.76  &  1.457 & & 0.135  & 18.03\\
  15.45  &  1.247 & & 0.160  & 17.94\\
  16.28  &  1.121 & & 0.292  & 17.72\\
  17.27  &  1.008 & & 0.517  & 17.70\\
  18.19  &  0.946 & &\nodata &\nodata\\
  19.06  &  0.902 & &\nodata &\nodata\\
  19.63  &  0.885 & &\nodata &\nodata\\
  19.86  &  0.884 & &\nodata &\nodata\\
  20.21  &  0.850 & &\nodata &\nodata\\
  20.45  &  0.773 & &\nodata &\nodata\\
  20.70  &  0.702 & &\nodata &\nodata\\
  20.90  &  0.675 & &\nodata &\nodata\\
  21.09  &  0.652 & &\nodata &\nodata\\
  21.29  &  0.648 & &\nodata &\nodata\\
  21.49  &  0.659 & &\nodata &\nodata \\
  21.80  &  0.676 & &\nodata &\nodata \\
  22.21  &  0.737 & &\nodata &\nodata \\
  22.60  &  0.808 & &\nodata &\nodata\\
\enddata
\end{deluxetable}

\end{document}